УДК 331.5

Табаков Валерій Зіновійович

*к.т.н., доц., доцент Інституту підготовки кадрів ДСЗУ*

# ІНДИКАТОРИ НАЯВНОСТІ НЕРИНКОВИХ ВІДНОСИН У СФЕРІ УПРАВЛІННЯ РИНКОМ ПРАЦІ УКРАЇНИ

*Виявлено факти, що свідчать про наявність неринкових відносин у сфері управління ринком праці України. Зроблено висновок про незаконне оподаткування коштів, сплачених легально працюючими в Україні, як страхових внесків на випадок безробіття. Зроблено висновок про посилення тиску з боку держави на органи регулювання ринку праці України, створеними на паритетній основі. Сформульовані рекомендації, щодо реалізації принципу вільної ринкової економіки в сфері регулювання ринку праці України.*

*Копенгагенські критерії, ЄС, принцип вільної ринкової економіки, регулювання ринку праці України.*

**Постановка проблеми**. 2014 рік є роком підписання Угоди про асоціацію України з Європейським Союзом (ЄС) [1]. 21 березня 2014 року прем'єр-міністр України Арсеній Яценюк від імені України підписав політичну частину Угоди. 27 червня 2014 року з моменту підписання Угоди президентом України Петром Порошенком, президентом Ради ЄС Германом Ван Ромпеєм, президентом Європейської комісії Жозе Мануелем Баррозу та головами держав та урядів усіх 27 країн-членів ЄС процес підписання Угоди між Україною та ЄС завершено. Угода передбачає політичну асоціацію та економічну інтеграцію України з ЄС. В разі ратифікації Угоди Верховною Радою України цей документ може бути тимчасово імплементований Україною без очікування ратифікації Угоди всіма 27 країнами-членами ЄС. Україна розглядає підписання Угоди про асоціацію з ЄС як важливий крок на шляху до членства України в Європейському Союзі. В преамбулі Угоди про асоціацію між Україною та ЄС підкреслюється, що подальші відносини між

Україною та ЄС базуватимуться на принципах вільної ринкової економіки, верховенстві права, ефективному урядуванні тощо.

З урахуванням зазначеного Україна має досягти відповідності схваленим на засіданні Європейської Ради [2], що відбулося у червні 1993 року в Копенгагені, критеріям, встановленим для країн-кандидатів на вступ до ЄС. Відповідно до Копенгагенських критеріїв, вимоги до членства у сфері економіки полягають «у наявності як дієвої ринкової економіки, так і здатності витримувати конкурентний тиск і дію ринкових сил у рамках ЄС».

За даних обставин предмет нашого дослідження, яке покликане визначити індикатори наявності неринкових відносин в сфері управління ринком праці України є актуальним.

**Аналіз останніх досліджень.** Ринок праці як окрема економічна система є предметом досліджень багатьох зарубіжних і українських вчених. Дж. Кейнс, А.Маршалл, П.Самуельсон, К.Макконелл, С.Брю, Р.Еренберг, Р.Сміт, В.Буланов, С.Карташов, Ю.Одегов, Л.Шевченко, С.Мочерний, О.Устенко сформували теоретичні основи регулювання ринку праці. Ринок праці як систему, що складається з економічної та інституціональної підсистем, досліджував Ю.Маршавін. Окремі питання теорії управління ринком праці і зокрема з позицій євроінтеграції України викладені в працях В.Федоренка, С.Бандура, Н.Діденко та О.Могильного. Практичні питання регулювання ринку праці України досліджували Д.Богиня, О.Грішнова, В.Васильченко, О.Карпіщенко, В.Петюх, О.Чернявська та ін.

**Невирішені частини проблеми та мета дослідження.** Проте внаслідок високого рівня специфічності здійснених наукових досліджень, а також високої швидкості зміни спрямованості соціальних запитів українського суспільства, сьогодні поза межами уваги зарубіжних та вітчизняних науковців залишаються питання наукового обґрунтування системи оцінки ступеня відповідності системи регулювання ринку праці України Копенгагенським критеріям. Специфіка нашого дослідження полягає спрямуванні вектора зусиль на виявлення та оцінку індикаторів наявності неринкових відносин у сфері управління ринком праці України, що дозволяє чітко визначити перелік необхідних управлінських рішень для якнайшвидшого досягнення системою регулювання ринком праці України рівня, що відповідає вимогам членства в ЄС.

**Матеріал та методика.** В нашому дослідженні по-перше звернемо увагу на оцінку поточної економічної ситуації в Україні Кабінетом Міністрів України (КМУ) і на характер рішень КМУ щодо управління діями інститутів системи регулювання ринком праці України. 1 березня 2014 року КМУ проаналізував стан економіки України, і зробив висновок, яким визнав економічну ситуацію в Україні нестабільною [3]. В Постанові від 1.03.2014 року №65 «Про економію державних коштів та недопущення втрат бюджету» КМУ розпорядився рекомендувати Правлінню Фонду загальнообов'язкового державного соціального страхування на випадок безробіття (ФЗДОССВБ) установити питому вагу видатків на організацію роботи (адміністративні, організаційно-управлінські, капітальні видатки, видатки на інформатизацію) в доходах без урахування залишку за попередній рік у розмірі 14,5% доходів Фонду та вжити заходів до економного та раціонального використання коштів, передбачених на утримання фондів, з урахуванням заходів, затверджених цією Постановою.

По-друге звернемось до законодавства України, що регулює відносини в сфері управління діями інститутів системи регулювання ринком праці України, і, зокрема, до ст. 10 Закону України «Про загальнообов'язкове державне соціальне страхування на випадок безробіття» від 02.03.2000 року [4], якою встановлено, що управління ФЗОДССВБ здійснюється на паритетній основі державою, представниками застрахованих осіб і роботодавців, і ст. 8 цього ж Закону, якою встановлено, що кошти Фонду не включаються до складу Державного бюджету України.

По-третє звернемо увагу на те, що згодом, починаючи приблизно з 2010 року, в Закон України «Про ЗОДССВБ» вносились зміни, що нівелювали згадані нами норми ст.ст. 8 та 10 Закону України «Про

ЗОДССВБ», в Закони України «Про Державний бюджет України»[5] обов'язково щорічно вносились норми про прирівнювання коштів ФЗОДССВБ до коштів Державного бюджету України, а в ч. 6 ст. 43 Бюджетного Кодексу України від 08.07.2010 року[6], що розміщена в главі 8 з назвою Казначейське обслуговування Бюджетних коштів, зазначено що під час здійснення операцій, пов'язаних з використанням коштів Фонду загальнообов'язкового державного соціального страхування України на випадок безробіття, застосовується казначейське обслуговування у порядку, визначеному Кабінетом Міністрів України.

Таким чином політика регулювання суспільних відносин у сфері ринку праці України поступово, але неухильно змінювала свій характер з ринкової, що здійснюється на паритетній основі до командної, управління в межах якої здійснюється прямими імперативними розпорядженнями держави в особі КМУ. Реакцією українського суспільства на такі зміни стало збільшення частки тіньової зайнятості та кількості трудових емігрантів.

Рух українського суспільства у напрямку до інтеграції з Європейським Союзом призвів до революційної зміни влади в Україні у 2014 році. Нова українська влада відгукнулася на суспільний запит декларацією беззастережного сприйняття європейських цінностей. Цей відгук дав надію на чітке дотримання українською владою базових принципів ЄС, а саме принципів вільних ринкових відносин, верховенства права, ефективного урядування, тощо, і недопущення застосування методів командної економіки в управлінні державою. В зв'язку з цим ми вирішили проаналізувати дії української влади у відомій нам сфері регулювання ринку праці України на предмет наявності або відсутності проявів застосування методів командного (неринкового) управління и тим самим знайти підтвердження або спростування того чи відповідають методи управління нової української влади базовим принципам ЄС.

Для здійснення нашого дослідження ми обрали метод порівняльного аналізу на підставі даних документу Постанови Правління ФЗОДССВБ від

29.05.2014 року № 10 «Про внесення змін до бюджету Фонду загальнообов'язкового державного соціального страхування України на випадок безробіття на 2014 рік»[7]. В зв'язку з утрудненим доступом до тексту цього документу наводимо його в даній публікації повністю:

ФОНД ЗАГАЛЬНООБОВ'ЯЗКОВОГО ДЕРЖАВНОГО СОЦІАЛЬНОГО СТРАХУВАННЯ УКРАЇНИ НА ВИПАДОК БЕЗРОБІТТЯ

ПРАВЛІННЯ ФОНДУ

ПОСТАНОВА

29.05.2014     м. Київ     №10

*Про внесення змін до бюджету Фонду загальнообов'язкового державного соціального страхування України на випадок безробіття на 2014 рік*

Відповідно до частини першої статті 11 Закону України „Про загальнообов'язкове державне соціальне страхування на випадок безробіття», пункту 14 Статуту Фонду загальнообов'язкового державного соціального страхування України на випадок безробіття (далі — Фонд), правління Фонду

**ПОСТАНОВЛЯЄ:**

1. Внести такі зміни до бюджету Фонду на 2014 рік, затвердженого постановою правління Фонду від 09.12.2013 №369:

1.1. Збільшити «Залишок коштів на початок року» на 378 179,0 тис. грн.;

1.2. Зменшити «Доходи поточного року — всього» на 3 666 466,6 тис. грн.;

1.3. Зменшити «Страхові внески» на 3 643 308,7 тис. грн.;

1.4. Зменшити «Інші надходження» на 21 256,8 тис. грн.;

1.5. Зменшити «Кошти Державного бюджету» на 1 901,1 тис. грн., в тому числі:

збільшити «на надання роботодавцям дотацій та компенсації для забезпечення молоді першим робочим місцем» на 1 039,9 тис. грн. та викласти в редакції «на надання роботодавцям компенсації для забезпечення молоді першим робочим місцем»;

зменшити «на соціальний захист працівників, що вивільняються у зв'язку з виведенням з експлуатації Чорнобильської АЕС» на 250,0 тис. грн.;

зменшити «на реєстрацію державною службою зайнятості трудових договорів, укладених між працівниками та фізичними особами» на 2 691,0 тис. грн.;

1.6. Зменшити «Всього доходів» на 3 288 287,6 тис. грн.;

1.7. Зменшити «Всього видатків» на 422 369,9 тис. грн.;

1.8. Зменшити видатки на «Матеріальне забезпечення та соціальні послуги — всього» на 466 979,5 тис. грн.;

1.9. Збільшити видатки на «Допомогу по безробіттю, в тому числі одноразову її виплату для організації безробітним підприємницької діяльності» на 1 193 879,8 тис. грн.;

1.10. Збільшити видатки на «Комісійну винагороду банкам за здійснення виплат матеріального забезпечення» на 3 581,4 тис. грн.;

1.11. Збільшити видатки на «Профілактику настання страхових випадків, допомогу по частковому безробіттю» на 5 167,7 тис. грн.;

L12. Зменшити видатки на «Здійснення заходів відповідно до Закону України «Про зайнятість населення» на 1 669 608,4 тис. грн.;

1.13. Зменшити видатки на «Розвиток та супроводження Єдиної інформаційно-аналітичної системи державної служби зайнятості» на 118 908,3 тис. грн.;

1.14. Збільшити видатки на «Відшкодування Пенсійному фонду витрат на виплату достроково призначеної пенсії» на 22 870,4 тис. грн.;

1.15. Зменшити видатки на «Утримання та забезпечення діяльності державної служби зайнятості, Інституту підготовки кадрів державної служби зайнятості, управління Фондом» на 134 250,6 тис. грн.;

1.16. Зменшити видатки на «Утримання державної служби зайнятості» на 132 258,2 тис. грн.;

1.17. Зменшити видатки на «Утримання Інституту підготовки кадрів державної служби зайнятості» на 792,4 тис. грн.;

1.18. Зменшити видатки на «НДР, методичні розробки та дослідження» на 1 000,0 тис. грн.;

1.19. Зменшити видатки на «Забезпечення міжнародного співробітництва» на 200,0 тис. грн.;

1.20. Зменшити видатки на «Створення умов прийому та надання соціальних послуг» на 97 844,3 тис. грн.;

1.21. Збільшити «Резерв Фонду» на 372 742,4 тис. грн.;

1.22. Зменшити «Залишок коштів на кінець року» на 2 865 917,7 тис. гривень.

2. Затвердити бюджет Фонду на 2014 рік із змінами, зазначеними у пункті 1 цієї постанови (додається).

2.1. Затвердити видатки на організацію роботи (на розвиток та супроводження Єдиної інформаційно-аналітичної системи державної служби зайнятості; утримання та забезпечення діяльності державної служби зайнятості, Інституту підготовки кадрів державної служби зайнятості, управління Фондом; створення умов прийому та надання соціальних послуг) у сумі 1 540 986,9 тис. гривень.

3. План реалізації у 2014 році Програми розвитку та супроводження Єдиної інформаційно-аналітичної системи державної служби зайнятості України на 2011-2014 роки, затверджений пунктом 2 постанови правління Фонду від 09.12.2013 №369, викласти у новій редакції, що додається.

4. План реалізації у 2014 році Програми створення робочими органами виконавчої дирекції Фонду умов прийому та обслуговування громадян та роботодавців на 2004-2016 роки, затверджений пунктом 3 постанови правління Фонду від 09.12.2013 №369, викласти у новій редакції, що додається.

5. Державному центру зайнятості:

5.1. здійснювати відшкодування витрат роботодавців на заробітну плату відповідно до договорів про забезпечення молоді першим робочим місцем з наданням роботодавцю дотації, що раніше укладені відповідно до Закону України «Про забезпечення молоді, яка отримала вищу або професійно-технічну освіту, першим робочим місцем з наданням дотації роботодавцю» за рахунок коштів в межах обсягів видатків за статтею «Працевлаштування безробітних шляхом надання дотацій роботодавцям»;

5.2. внести зміни, згідно цієї постанови, до зведеного кошторису видатків Фонду загальнообов'язкового державного соціального страхування України на випадок безробіття на 2014 рік.

5.3. Підготувати лист за підписом Голови правління до Кабінету Міністрів України та Міністерства соціальної політики України про стан виконання рекомендацій, передбачених пунктом 3 постанови Кабінету Міністрів України від 01.03.2014 р. № 65 «Про економію державних коштів та недопущення втрат бюджету».

Голова правління Фонду                                              С.Кондрюк

В наслідок прийняття Правлінням ФЗОДССВБ 29.05.2014 року Постанови №10 Бюджет ФЗОДССВБ на 2014 рік став таким [табл.1]:

Таблиця 1.

**Бюджет ФЗОДССВБ на 2014 рік**

| | Статті | тис.грн. |
|---|---|---|
| **1.** | **Залишок коштів на початок року** | **4 619 718,8** |
| **2.** | **Доходів поточного року – всього:** | **6 448 774,2** |
| | з них: | |
| 2.1. | Страхові внески | 6 403 039,8 |
| 2.2 | Інші надходження | 36 381,3 |
| 2.3 | Кошти Державного бюджету | 9 353,1 |
| | З них: | |
| | На надання роботодавцям компенсації для забезпечення молоді першим робочим місцем | 8 950,0 |
| | На соціальний захист працівників, що вивільняються у зв'язку з виведенням з експлуатації Чорнобильської АЕС. | 403,1 |

| 3. | **Всього доходів** (разом із залишком коштів на початок року) | **11 068 493,0** |
|---|---|---|
| 4. | **Всього видатків** (разом з резервом) | **11 068 493,0** |
| 4.1 | **Матеріальне забезпечення та соціальні послуги – всього:** | **8 570 212,7** |
| 4.1.1. | Допомога по безробіттю, у тому числі одноразова її виплата для організації безробітним підприємницької діяльності | 6 938 040,6 |
| 4.1.2. | Допомога на поховання | 2 312,9 |
| 4.1.3. | Комісійна винагорода банкам за здійснення виплат матеріального забезпечення | 20 867,0 |
| 4.1.4. | Професійна підготовка, перепідготовка та підвищення кваліфікації | 546 572,5 |
| 4.1.5. | Видача ваучерів громадянам старше 45 років. | 54 539,8 |
| 4.1.6. | Організація громадських робіт | 55 404,8 |
| 4.1.7. | Працевлаштування безробітних шляхом надання дотацій роботодавцям | 21 490,4 |
| 4.1.8. | Компенсація єдиного внеску роботодавцям | 130 362,1 |
| 4.1.9. | Інформаційні та консультаційні послуги, пов'язані з працевлаштуванням та їх забезпечення | 120 155,6 |
| 4.1.10. | Профорієнтація та її забезпечення | 11 247,5 |
| 4.1.11. | Профілактика настання страхових випадків, допомога по частковому безробіттю | 11 267,9 |
| 4.1.12. | Здійснення заходів відповідно до Закону України «Про зайнятість населення» | 657 951,6 |
| 4.2. | **Розвиток та супроводження Єдиної інформаційно-аналітичної системи державної служби зайнятості** | **121 873,6** |
| 4.3. | **Відшкодування Пенсійному фонду витрат на виплату достроково призначеної пенсії** | **105 870,9** |
| 4.4. | **Утримання та забезпечення діяльності державної служби зайнятості, Інституту підготовки кадрів державної служби зайнятості, управління Фондом** | **1 321 957,6** |
| 4.4.1. | Утримання державної служби зайнятості | 1 295 931,0 |
| 4.4.2. | Утримання Інституту підготовки кадрів державної служби зайнятості | 24 672,1 |
| 4.4.3. | Забезпечення поточної діяльності правління Фонду | 220,0 |
| 4.4.4. | Підвищення кваліфікації працівників державної служби зайнятості | 884,5 |
| 4.4.5. | НДР, методичні розробки та дослідження | 0,0 |
| 4.4.6. | Забезпечення міжнародного співробітництва | 250,0 |
| 4.5. | **Створення умов прийому та надання соціальних послуг** | **97 155,7** |
| 4.6. | **Резерв Фонду** | **851 422,5** |
| 5. | **Залишок коштів на кінець року** | **0,0** |

Перейдемо до аналізу даних цього документу. Для цього складемо порівняльну таблицю [табл.2] основних статей Бюджету ФЗОДССВБ: 1. Залишок на кінець 2013 року. 2. Розмір Резерву Фонду на кінець 2013 року. 3. Доходи Фонду в 2014 році. 4. Видатки Фонду в 2014 році. 5. Розмір Резерву Фонду на кінець 2014 року. 6. Залишок Фонду на кінець 2014 року, до прийняття Постанови від 29.05.2014 та після прийняття Постанови від. 29.05.2014.

Таблиця 2

**Порівняльна таблиця основних статей Бюджету ФЗОДССВБ на 2014 рік**

| Назва | Розмір до прийняття Постанови 29.05.2014 (тис. грн.) | Розмір після прийняття Постанови 29.05.2014 (тис. грн.) |
|---|---|---|
| Залишок на кінець 2013 року. | 4241539,8 | 4619718,8 |
| Розмір Резерву Фонду на кінець 2013 року | 537546,0 | 537546,0 |
| Розмір допомоги по безробіттю, у тому числі одноразової її виплати для організації безробітними підприємницької діяльності | 5744160,8 | 6938040,6 |
| 1/12 розміру допомоги по безробіттю, у тому числі одноразової її виплати для організації безробітними підприємницької діяльності[*] | **478680,1** | **578170,6** |
| Доходи Фонду в 2014 році. | 10115240,8 | 6448774,2 |
| Видатки Фонду в 2014 році | 11012182,8 | 10217070,5 |
| Розмір Резерву Фонду на кінець 2014 року. | **478680,1** | **851422,5** |
| Залишок Фонду на кінець 2014 року | 2865917,7 | 0 |

[*] Введено для порівняння.

З таблиці 2 бачимо: після прийняття Постанови Правління Фонду ЗОДССВБ 29.05.2014 року плановий залишок Фонду на початок 2014 (кінець 2013) року зріс на 400 млн. грн., заплановані доходи Фонду в 2014 році зменшилися на 4 млрд. грн., заплановані видатки Фонду у 2014 році зменшилися на 0,8 млрд. грн., залишок Фонду на кінець 2014 року зведений до нуля, резерв Фонду в 2014 році збільшено на 400 млн. грн.

Для забезпечення надійності функціонування системи регулювання ринку праці України в бюджеті ФЗОДССВБ утворюється Резерв Фонду. Зазначений резерв для системи, яка є самодостатньою має відповідати визначеній частці загальному розміру видатків Фонду. Оскільки видатки Фонду зменшені Постановою від 29.05.2014 року на 10%, то за логікою забезпечення системи регулювання ринку праці розмір Резерву Фонду слід було б відповідно зменшити на 10%. Але логіка забезпечення функціонування системи регулювання ринку праці України шляхом резервування коштів Фонду дещо інша. Згідно з п. 43 Статуту ФЗОДССВБ [8] резерв Фонду формується шляхом щомісячного відрахування до нього 1/12 частини коштів, призначених в поточному місяці на виплату допомоги по безробіттю в поточному місяці. Тому ми в нашому дослідженні вирахували 1/12 розміру допомоги по безробіттю, у тому числі одноразової її виплати для організації безробітними підприємницької діяльності та порівняли її з розміром Резерву Фонду. До прийняття Постанови від 29.05.2014 року розмір Резерву Фонду повністю відповідає 1/12 розміру допомоги по безробіттю і становить 478 680,1 тис. грн. Після прийняття Постанови від 29.05.2014 року розмір Резерву Фонду складає 851 422,5 тис. грн., а 1/12 розміру допомоги по безробіттю становить 578 170,6 тис. грн. Таким чином в результаті прийняття Правлінням ФЗОДССВБ Постанови від 29.05.2014 року Резерв Фонду всупереч вимог п. 43 Статуту ФЗОДССВБ визначено з перебільшенням на суму 273 251,9 тис. грн. Оскільки Постановою від 29.05.2014 року не визначені причини такої невідповідності вимогам Статуту Фонду та статті витрат утворюваного надлишку резервних коштів Фонду, дана обставина свідчить про можливі наміри використовувати утворювані надлишки не за прямим призначенням коштів Фонду, тобто не на боротьбу з безробіттям.

Взагалі внаслідок дії механізму банківської мультиплікації в умовах жорсткого контролю за діями Правління ФЗОДССВБ, як розпорядника коштів Фонду, з боку Держказначейства, а також імперативної вимоги КМУ про зменшення видатків Фонду, створює на нашу думку ефект подвійного та кратного утримання податку з коштів Фонду, які були отримані в результаті сплати легально працюючими в Україні обов'язкового страхового внеску на випадок безробіття.

**Висновки та перспективи подальших досліджень.** Таким чином на підставі проведеного аналізу ми можемо зробити ряд висновків:

1. Характер дій української влади в умовах декларування нею беззастережного сприйняття європейських цінностей та дотримання базових принципів ЄС, таких як принципи вільної ринкової економіки, верховенства права, ефективного урядування тощо, не відрізняється від характеру дій української влади до революційних подій лютого 2014 року. Замість впровадження принципу вільної ринкової економіки у сфері регулювання ринку праці України спостерігається прямий імперативний диктат з боку КМУ по відношенню до органів регулювання ринку праці України, утворених на паритетній основі.

2. Імперативний диктат з боку КМУ по відношенню до органів регулювання ринку праці України, утворених на паритетній основі, викликаний узурпацією з боку КМУ повноважень Правління ФЗОДССВБ щодо розпорядженням коштами Фонду ЗОДССВБ. Даний диктат має характер подвійного та кратного оподаткування коштів легально працюючих в Україні, які отримані як податок (страховий внесок на випадок безробіття). Метою даного диктату є бажання Кабінету міністрів України використовувати кошти Резерву ФЗОДССВБ в умовах запровадження жорсткого контролю за діями Правління ФЗОДССВБ, як розпорядника коштів Фонду, з боку Держказначейства, не за призначенням, тобто не на потреби боротьби з безробіттям, що стає можливим в результаті дії механізму банківської мультиплікації.

3. Вимога КМУ, сформульована в Постанові від 1.03.2014 року №65 «Про економію державних коштів та недопущення втрат бюджету» про рекомендацію Правлінню ФЗДОССВБ установити питому вагу видатків на організацію роботи (адміністративні, організаційно-управлінські, капітальні видатки, видатки на інформатизацію) в доходах без урахування залишку за попередній рік у розмірі 14,5% доходів Фонду та вжити заходів до економного та раціонального використання коштів, передбачених на утримання фондів, з урахуванням заходів, затверджених цією Постановою, є прямим втручанням держави в діяльність органу регулювання ринком праці України, створеним на паритетній основі, і кваліфікується нами як індикатор наявності неринкових відносин в сфері регулювання ринком праці України.

4. Утворення надлишку коштів Резерву ФЗОДССВБ в результаті ухвалення Правлінням ФЗОДССВБ Постанови від 29.05.2014 року №10 «Про внесення змін до бюджету Фонду загальнообов'язкового державного соціального страхування України на випадок безробіття на 2014 рік» за вимогою Постанови КМУ від 01.03.2014 року №65 «Про економію державних коштів та недопущення втрат бюджету» свідчить про посилення пресу командного стилю управління шляхом видання КМУ прямих рекомендацій Правлінню Фонду ЗОДССВБ, як органу регулювання ринку праці України, утвореним на паритетній основі. Таким чином декларування новою українською владою беззастережного сприйняття європейських цінностей і зокрема принципу вільної ринкової економіки не підтверджується фактичними діями нової української влади. Дії нової української влади по відношенню до органів регулювання ринку праці України, створених на паритетній основі свідчать про посилення диктату нової української влади порівняно з аналогічним диктатом української влади, що діяла до подій 18 лютого 2014 року.

Для реалізації принципів вільної ринкової економіки в Україні на нашу думку може бути рекомендовано:

1. Виключити з українського законодавства норми, що забезпечують жорсткий контроль за діями Правління ФЗОДССВБ, як розпорядника коштів Фонду, з боку Держказначейства.

2. Забезпечити повне і беззастережне управління коштами ФЗОДССВБ Правлінням Фонду, як органом, що створений на паритетній основі державою, представниками застрахованих осіб і роботодавців.

3. Виключити можливість використання коштів ФЗОДССВБ як коштів Державного бюджету України.

Табаков Валерий Зиновьевич, **Индикаторы наличия нерыночных отношений в сфере управления рынком труда Украины.**

*Выявлены факты, свидетельствующие о наличии нерыночных отношений в сфере управления рынком труда Украины. Сделан вывод о незаконном налогообложения средств, уплаченных легально работающими в Украине, как страховых взносов на случай безработицы. Сделан вывод об усилении давления со стороны государства на органы регулирования рынка труда Украины, созданными на паритетной основе. Сформулированы рекомендации по реализации принципа свободной рыночной экономики в сфере регулирования рынка труда Украины.*

*Копенгагенские критерии, ЕС, принцип свободной рыночной экономики, регулирование рынка труда Украины.*

Tabakov Valery Zinovievich. **Indicators of availability of non-market relations in the sphere of labor market in Ukraine.**

*There are identified indicators of availability a non-market relations in the sphere of labor market in Ukraine. It is concluded that illegal tax money paid by*

*legally working in Ukraine, as insurance premiums in the event of unemployment. It is concluded that increased pressure from the government on labor market regulators Ukraine established on a parity basis. There are formulated recommendations for the implementation of the principle of a free market economy in the regulation of the labor market of Ukraine.*

*Copenhagen criteria, the EU, the principle of a free market economy, labor market regulation in Ukraine.*